\documentclass[12pt]{amsart}

\newcommand {\supplus}{\mathop{{\supset}\llap{\raise 
0.5pt\hbox{\normalfont\small+}\hskip 0.5pt}}} 

\newcommand {\subplus}{\mathop{{\subset}\llap{\raise 
0.5pt\hbox{\normalfont\small+}\hskip 0.5pt}}}  

\newcommand {\Cee}    {{\mathbb  C}}

\newcommand {\Kee}    {{\mathbb  K}}

\newcommand {\Ree}    {{\mathbb  R}}

\newcommand {\Zee}    {{\mathbb  Z}}

\newcommand {\fder}   {{\mathfrak{der}}}   %

\newcommand {\fg}     {{\mathfrak{g}}}    %
\newcommand {\fgl}    {{\mathfrak{gl}}}  %
\newcommand {\fk}     {{\mathfrak{k}}}

\newcommand {\fm}     {{\mathfrak{m}}}

\newcommand {\fns}     {{\mathfrak{ns}}}

\newcommand {\fosp}   {{\mathfrak{osp}}}

\newcommand {\fr}     {{\mathfrak{r}}}

\newcommand {\fsl}    {{\mathfrak{sl}}}

\newcommand {\fst}    {{\mathfrak{st}}}
\newcommand {\fsvect} {{\mathfrak{svect}}}

\newcommand {\fvect}  {{\mathfrak{vect}}}   %
\newcommand {\fvir}   {{\mathfrak{vir}}}

\newcommand {\fwitt}  {{\mathfrak{witt}}}

\newcommand {\fz}     {{\mathfrak{z}}}

\newcommand {\cal} {\mathcal}

\newcommand {\cC}     {{\cal C}}
\newcommand {\cD}     {{\cal D}}

\newcommand {\cF}     {{\cal F}}

\newcommand {\cK}     {{\cal K}}
\newcommand {\cL}     {{\cal L}}

\newcommand {\cT}     {{\cal T}}

\def \opname#1#2%
  {\expandafter\newcommand \csname #1\endcsname {{\mathop{#2}\nolimits}}}

\newcommand{\rmname}[1]
  {\expandafter\newcommand \csname #1\endcsname {{\operatorname{#1}}}}

\newcommand{\rmnameii}[2]
  {\expandafter\newcommand \csname #1\endcsname {{\operatorname{#2}}}}

\rmname{act}
\rmname{Ad}
\rmname{Add}
\rmname{ad}
\rmname{Alt}
\rmname{alt}
\rmname{Ann}
\rmname{antidiag}
\rmname{Ber}
\rmname{ber}
\rmname{Br}
\rmname{card}
\rmname{ch}
\rmname{Char}
\rmname{cem}
\rmname{cj}
\rmname{Cliff}
\rmname{cntr}
\rmname{codim}
\rmname{coind}
\rmname{const}
\rmname{col}
\rmname{cork}
\rmname{cpr}
\rmname{diag}
\rmnameii{Div}{div}
\rmname{Def}
\rmname{Der}
\rmname{Dim}
\rmname{End}
\rmname{Even}
\rmname{Ext}
\rmname{gr}
\rmname{Hom}
\rmname{HT}
\rmnameii{Ht}{ht}
\rmname{hwt}
\rmname{Id}
\rmname{id}
\rmname{ind}
\rmname{Ind}
\rmname{Inf}
\rmname{irr}
\rmname{Le}
\rmname{Lie}
\rmname{lwt}
\rmname{mult}
\rmname{Mor}
\rmname{nm}
\rmname{Ob}
\rmname{Odd}
\rmname{Osc}
\rmname{per}
\rmname{Pic}
\rmname{pr}
\rmname{pro}
\rmname{Prime}
\rmname{Proj}
\rmname{prt}
\rmname{pt}
\rmname{Q}
\rmname{qet}
\rmname{qtr}
\rmname{rd}
\rmname{rk}
\rmname{row}
\rmname{Res}
\rmname{salt}
\rmname{Sch}
\rmname{SBr}
\rmname{scalar}
\rmname{Ser}
\rmname{sign}
\rmname{Smbl}
\rmname{spin}
\rmname{ssym}
\rmname{str}
\rmname{st}
\rmname{sgn}
\rmname{sq}
\rmname{symm}
\rmname{supp}
\rmname{Supp}
\rmname{St}
\rmname{Spec}
\rmname{Spm}
\rmname{tr}
\rmname{vpt}
\rmname{weyl}
\rmname{Weyl}
\rmname{Witt}

\opname{vvol}  {{v\hspace{-0.1ex}o\hspace{-0.02ex}l\/}}
\opname{pnt}  {\text{\normalfont pt}}
\opname{Span} {{Span}}
\opname{slim} {\overline{\lim}}
\opname{Vol}  {{V\hspace{-0.55ex}o\hspace{-0.02ex}l\/}}
\opname{Par}  {{P\hspace{-0.3ex}a\hspace{-0.05ex}r\/}}

\rmname{Mat}
\rmname{Bil}
\rmname{Diff}
\rmname{Ker}
\rmname{Herm}
\rmname{Coker}
\rmname{Conn}
\rmname{Covect}
\rmname{Vect}
\rmname{Int}

\newcommand {\ev} {{\bar0}}
\newcommand {\od} {{\bar1}}

\newcommand {\tto} {\longrightarrow}
\newcommand {\pder}[1] {{\frac{\partial}{\partial {#1}}}}
\newcommand {\pderf}[2] {{\frac{\partial {#1}}{\partial {#2}}}}

\newcommand {\bcdot}   {\mathbin{\hbox{\raise.4ex\hbox{\bf.}}}} 

\newcommand {\secno} {}

\newtheorem{Theorem}{\secno Theorem}

\newenvironment {th*}[1]
    {\gdef\thname{#1} \begin{thn}}%
    {\end{thn}}
\newtheorem{thn}[Theorem] {\thname}

\theoremstyle{definition}

\newenvironment {ex*}[1]
    {\gdef\thname{#1} \begin{exn}}%
    {\end{exn}}
\newtheorem{exn}[Theorem]{\thname}

\theoremstyle{remark}

\newenvironment {rem*}[1]
    {\gdef\thname{#1} \begin{remn}}%
    {\end{remn}}
\newtheorem{remn}[Theorem]{\thname}

\newcommand {\ssec}{\subsection*}

\begin{document}

\title[Schr\"odinger and
Korteweg--de Vries operators]{Supersymmetry of the Schr\"odinger and
Korteweg--de Vries operators}

\author{D.~Leites${}^1$, Xuan P.${}^2$}

\address{${}^1$ (Address for correspondence) Dept. of Math.,
Univ. of Stockholm, Roslagsv. 101,
Kr\"aftriket hus 6, S-106 91, Stockholm, Sweden;  ${}^2$ Dept. of Phys., Univ. of
Berkeley, USA}

\thanks{Financial support of the
NFR is gratefully acknowledged.}



\begin{abstract} In 70's A.A.~Kirillov  
interpreted the (stationary) {\it Schr\"odinger} or {\it 
Sturm-Liouville} operator $L_2=\frac{d^2}{dx^2}+ F$ as an element of 
the dual space $\hat\fg^{*}$ to 
$\hat\fg=\fvir$, the Virasoro algebra; he also
interpreted the (stationary) {\it KdV operator} 
$L_3=\frac{d^3}{dx^3}+\frac{d}{dx}F+ F\frac{d}{dx}$ in terms of the 
stabilizer of $L_2$ and found a supersymmetry that connects 
solutions of $L_3f=0$ with solutions of $L_2g=0$.

We extend Kirillov's results and find all supersymmetric extension of 
the Schr\"odinger and Korteweg--de Vries operators.  

We also superize the construction Khesin--Malikov's generalization of 
Drinfeld--Sokolov's reduction to the pseudodifferential operators.  
\end{abstract}

\maketitle
\hskip 7 cm  To Alexandr Alexandrovich Kirillov

\hskip 7 cm  who taught one of us
\vskip 0.2 cm

\section*{Introduction} In 70's A.A.~Kirillov made several amazing 
observations.  In [Ki1] he associated the (stationary) {\it KdV 
operator} $L_3=\frac{d^3}{dx^3}+\frac{d}{dx}F+ F\frac{d}{dx}$ and the 
(stationary) {\it Schr\"odinger} or {\it Sturm-Liouville} operator 
$L_2=\frac{d^2}{dx^2}+ F$ with the cocycle that determines the 
nontrivial central extension $\hat\fg=\fvir$ --- the Virasoro algebra 
--- of the Witt algebra $\fg=\fwitt=\fder\Cee[x^{-1}, x]$.  Moreover, 
he found an explanation of the commonly known fact that the product of 
two solutions $f_1$, $f_2$ of the Schr\"odinger equation $L_2(f)=0$ 
satisfies $L_3(f_1f_2)=0$.  Kirillov's explanation: a supersymmetry 
[Ki2].

Kirillov also classified the orbits of the coadjoint representation of
$\fvir$ and clarified its equivalence to the following important classification
problems: of symplectic leaves of the second Gelfand--Dickey structure on the
2nd order differential operators, of projective structures on the circle and of
Hill equation (i.e., the Schr\"odinger equation with periodic potential).

Kirillov's approach clarifies the earlier results by Poincar\'e, N. Kuiper,
Lazutkin and Pankratova. The recent announcement of the classification of the
simple stringy superalgebras [GLS] describes, clearly,  the scope of the problem:
there are exactly 14 ways to superize the above result of Kirillov. (Kirillov himself
partly considered one of them; several first  few of the 14 possibilities were considered
by V. Ovsienko and O. Ovsienko, Radul, Khesin and others, see [L2], [KM], [BIK],
[DI], [DIK].)

We extend Kirillov's results and find all supersymmetric extension of 
the Schr\"o\-din\-ger and Korteweg--de Vries operators associated with the 
10 distinguished stringy superalgebras, i.e., all the simple stringy 
superalgebras that possess a nontrivial central extension.  

We also superize the construction due to Khesin--Malikov  
(Drinfeld--Sokolov's reduction for the pseudodifferential operators) 
and relate the complex powers of the Schr\"o\-din\-ger operators we describe 
to the superized KdV-type hierarchies labelled by complex parameter.  

Our construction brings the KdV-type equations directly in the Lax 
form guaranteeing their complete integrability.  The examples of 
$N$-extended KdV-type equations due to Chaichian--Kulish, 
Kuperschmidt, P. Mathieu, Ivanov--Krivonos--Bellucci--Delduc and 
others are only connected with several of the distinguished stringy 
superalgebras and not all of the cocycles we consider.

Our superizations of KdV are completely integrable by construction.  
On the other hand, physicists can construct by a different method {\it 
parametric} families of KdV-type equations on the supercircle which 
are completely integrable for some values of the parameter only.  We 
can only interprete some particular of these equations.

\ssec{0.1. Kirillov's interpretation of the Schr\"odinger and
Korteweg--de Vries operators} Let $\fg=\fwitt=\fder\Cee[x^{-1}, x]$; let
$\hat\fg=\fvir$ be the nontrivial central extension of $\fg$ given by the
cocycle
$$
[f\frac{d}{dx}+az, g\frac{d}{dx}+bz]=(fg'-f'g)\frac{d}{dx}+c\cdot \Res
fg'''\cdot z\; \text{for } c\in \Cee,
$$
where $z$ is the generator of the center of $\hat\fg$.  Let $\cF 
=\Cee[x^{-1}, x]$ be the algebra of functions; let $\cF _{\lambda}$ be 
the rank 1 module over $\fg$ and $\cF$ spanned over $\cF $ by 
$dx^{\lambda}$, where the $\lambda$th power of $dx$ is determined via 
analiticity of the formula for the $\fg$-action:
$$
(f\frac{d}{dx})(dx^{\lambda})=\lambda f'dx^{\lambda}.
$$
In particular, $\fg\cong\cF _{-1}$, as
$\fg$-modules. Since the module $\Vol$ of volume forms is $\cF _1$, the module
dual to $\fg$ is
$\fg^*=\cF _{2}$: we use one $dx$ to kill $\frac{d}{dx}$ and another $dx$ to
integrate the product of functions. (We confine ourselves to {\it regular}
generalized functions, i.e., we ignore the elements from the space of
functionals on
$\fg$ with 0-dimensional support, see [Ki1].) Explicitely,
$$
F(dx)^2(f\frac{d}{dx})=\Res~Ff.
$$

$\bullet$ The Lie algebra of the stationary group of the element $\hat
F=(F, c)\in
\hat\fg^*=(\fg^*, \Ree\cdot z^*)$ is
$$
\fst_{\hat F}=\{\hat X\in \hat\fg: \hat F([\hat X, \hat Y])=0\;
\text{for any }\hat Y\in \hat\fg\}.
$$
Take $\hat X=g\frac{d}{dx}+az$, $\hat Y=f\frac{d}{dx}+bz$. Then
$$
\begin{matrix}
\hat F([\hat X, \hat Y])=\widehat F{[(fg'-f'g)\frac{d}{dx}+\Res
fg'''\cdot z]}=\\
\Res [F(fg'-f'g)+cfg''']\overset{(\text{partial integration})}{=}\Res
f[Fg'+(Fg)'+cg'''].\end{matrix}
$$
Hence, $\hat X\in \fst_{\hat F}$ if and only if $g$ is a solution of
the equation $L_3g=0$.
If $c\neq 0$ we can always rescale the equation and assume that
$$
c=1.\eqno{(0.1.1)}
$$
In what follows this is understood. We call $L_3$ the {\it KdV operator}.
We also see
that the KdV operator can be expressed in the form
$$
L_3=(\text{the cocycle operator that determines $\hat\fg$
})+\frac{d}{dx}F+F\frac{d}{dx}.\eqno{(0.1.2)}
$$

$\bullet$ Observe that the transformation properties of $z^*$ and
$(dx)^2\frac{d^2}{dx^2}$ under $\fg$ are identical: scalars. The
Schr\"odinger operator
$L_2=\frac{d^2}{dx^2}+ F$ is, clearly, selfadjoint. The factorization
$$
F(dx)^2+a\cdot z^*=(dx)^2(F+ a\frac{d^2}{dx^2})\eqno{(0.1.3)}
$$
suggests to represent the elements of $\hat\fg^*$ as 2nd order selfadjoint
differential operators: $\cF_\lambda\tto \cF_{\lambda+2}$. The
selfadjointness fixes $\lambda$; indeed,
$1-(\lambda+2)=\lambda$, i.e., $\lambda=-\frac{1}{2}$.

$\bullet$ Assume that $F$ depends on time, $t$ is the corresponding
parameter. The
{\it KdV hierarchy} is the series of evolution equations for $L=L_2$ or,
equivalently, for $F$ :
$$
\Dot L=[L, A_k], \text{ where } A_k=(\sqrt{L}^{2k-1})_+ \text{ for } k= 1,
3, 5, \dots .\eqno{(0.1.4:\text{KdV$_k$ hierarchy})}
$$
Here the subscript $+$ singles out the differential part of the
pseudodifferential operator. The case $k=1$ is trivial and $k=3$ corresponds
to the original KdV equation.

$\bullet$ Khesin and Malikov ([KM]) observed that we can also consider evolution
equations for PDO:
$$
\Dot L=[L, A_\lambda], \text{ where } A_\lambda=\left (\sqrt{L}\right)^{\lambda}
\text{ for }
\lambda\in\Cee.\eqno{(0.1.5)}
$$
Such an approach to evolution equations for $L_2$ is, as we will see, even
more natural
in the supersetting, when the Schr\"odinger operator becomes a
pseudodifferential
one itself.

\ssec{0.2. Kirillov's interpretation of supersymmetry of the Schr\"odinger and
Korteweg--de Vries operators} (To better understand this subsection,
the reader has to know the technique of
$C$-points, see [L1], [L2] and \S 1.)  If in the above scheme we
replace
$\fg=\fwitt$ with the Lie superalgebra
$\fg=\fk^L(1|1)$ of contact vector fields on the $1|1$-dimensional supercircle
associated with the trivial bundle, we get two operators (for the definition of
$K_f$ see sec. 1.4)
$$
\cL_5=K_\theta K_1^2+2FK_1+2K_1F+(-1)^{p(F)}K_\theta FK_\theta \eqno{(0.2.1:\text{the
$\fns$ analog of $L_3$, the KdV operator})}
$$
and
$$
\cL_3=K_\theta K_1+F.\eqno{(0.2.2:\text{the $\fns$
analog of $L_2$, the Schr\"odinger operator})}
$$
Here $F\in\Pi(\Cee[x^{-1}, x, \theta])$ and $K_f$ is the contact vector field
generated by $f\in\Cee[x^{-1}, x, \theta]$.

Indeed, let us calculate the stabilizer of an element of $\fns(1)^*$. In doing
so we will use the {\it $C$-points} of all objects encountered (physicists
call such objects {\it superfields}). 

Observe that since the integral (or residue) pairs $1$ with
$\frac{\theta}{x}$, this pairing is odd, and, therefore, $\fns(1)^*=\Pi(\cF
_3)$.

The straightforward calculations yield:
$X=K_f\in \fst_{\hat F}$ if and only if $f$ is a solution of
the equation
$$
\left(cK_\theta\frac{d^2}{dx^2}+2\frac{d}{dx}F+
2F\frac{d}{dx}+(-1)^{p(F)}K_\theta FK_\theta \right)f=0.
\eqno{(0.2.3):\text{KdV}(\fns(1))}
$$
The operator
$$
\cL_5=(\text{the cocycle operator that determines
$\hat\fg$})+2FK_1+2K_1F+(-1)^{p(F)}K_\theta K_1K_\theta\eqno{(0.2.4)}
$$
from the lhs of (0.2.3) will be called the
$\fns(1)$-{\it KdV operator}.

In components we have: $f=f_0+f_1\theta$, $F=F_0+F_1\theta$, where
$f_0$ and $F_1$ are even functions (of $t$ with values in an auxiliary
supercommutative superalgebra $C$) while
$f_1$ and
$F_0$ are odd ones. Equation $(\text{KdV}(\fns(1)))$ takes the form:
$$
\left [ \begin{pmatrix}L_3&0\\
     0&L_2\end{pmatrix}+ \begin{pmatrix}
0&2F_0\frac{d}{dx}+\frac{d}{dx}F_0\\
F_0\frac{d}{dx}+2\frac{d}{dx}F_0&0 \end{pmatrix} \right] \begin{pmatrix}
f_0\\
f_1\end{pmatrix}=0. \eqno{(0.2.5):\left(\begin{matrix}\text{KdV}(\fns(1))\\
\text{in components}\end{matrix}\right)}
$$

Suppose $F_0=0$. Since formula (1.4.5) implies that $\{f(t)\theta,
g(t)\theta\}_{K.b.}=fg$, we see that the product of two solutions of the
Schr\"odinger equation $L_2f=0$ is a solution of the
KdV equation $L_3X=0$. This is Kirillov's supersymmetry.

\begin{rem*}{Remark} Kirillov only considered $\Ree$-points of $\fg$ 
and $\hat\fg$, that is why he missed all odd parameters of the supersymmetry he 
found --- the second summand of the KdV operator
$$
\cL_5= \begin{pmatrix}L_3&0\\
     0&L_2\end{pmatrix}+ \begin{pmatrix}
0&2F_0\frac{d}{dx}+\frac{d}{dx}F_0\\
F_0\frac{d}{dx}+2\frac{d}{dx}F_0&0 \end{pmatrix}.\eqno{(0.2.6)}
$$
\end{rem*}

\ssec{0.3.  Our result} We extend Kirillov's result 
from $\fwitt$ to all simple distinguished stringy Lie superalgebras.  
We thus elaborate the remark from [L1], p.  167, where the importance 
of odd parameters in this problem was first observed and the problem 
solved here was raised.  To consider {\it all} superized KdV and 
Schr\"odinger operators was impossible before the list of stringy 
superalgebras and their cocycles (see [GLS]) was completed.

Passing to superization of the bulleted steps of sec. 0.1, we have to consider
the elements of $\hat\fg^*$ for the distinguished stringy superalgebras $\fg$
as selfadjoint operators, perhaps, pseudodifferential, rather than
differential. This together with ideas applied by Khesin--Malikov to the
usual Schr\"odinger operator requires generalizations of the Lie superalgebra
of matrices of complex size associated with the analogs of superprincipal
embeddings of $\fosp(N|2)$ for $N\leq 4$. Such superizations were recently
described together with a description of the corresponding
$W$-superalgebras and Gelfand--Dickey superalgebras, 
see [GL].

There remains a puzzle. In physical papers ([BIK], [DI], [DIK] and refs.
therein) {\it parametric} families of KdV-type equations
on the supercircle are introduced. The equations are completely integrable for
some values of the parameter only. We can only identify some particular of
these equations. How to describe the equations for the other values of
parameter in terms of the supersymmetry algebra?

\section*{\S 1. Distinguished stringy superalgebras}

We recall all the neccessary data. For the details of classification of
simple finite
dimensional Lie superalgebras see [K1], [LSch] and [GLS]; for a review of the
representation theory of simple Lie superalgebras including infinite
dimensional ones
see [L2], for basics on supermanifolds see [L1] or [M]. The ground field is
$\Cee$.

\ssec{1.1. Supercircle} A {\it supercircle} or (for a physicist) a {\it
superstring}
of dimension $1|n$ is the real supermanifold $S^{1|n}$ associated with the rank
$n$ trivial vector bundle over the circle. Let $x=e^{i\varphi}$, where
$\varphi$ is the angle parameter on the circle, be the even indeterminate of the
Fourier transforms; let $\theta=(\theta_1, \dots, \theta_n)$, be the odd
coordinates
on the supercircle formed by a basis of the fiber of the trivial bundle over the
circle. Then $(x, \theta)$ are the coordinates on $(\Cee^*)^{1|n}$, the
complexification of
$S^{1|n}$.

Denote by $\vvol=\vvol(x,\theta)$ the volume element on $(\Cee^*)^{1|n}$.
(Roughly
speaking, $\vvol\lq\lq="\frac{dx}{d\theta_1\dots d\theta_n}$, or even
better $\vvol\lq\lq="du_1\cdot ...\cdot du_m\cdot\pder{\xi_1}\cdot...\cdot
\pder{\xi_n}$. Recall that, actually, these are not equalities: as shown in
[BL], the change of variables acts differently on the lhs and rhs and only
coinsides for the simplest transformations.)

Let the contact form be
$$
\alpha= dx -\sum_{1\leq i\leq n}\theta_id\theta_i.
$$
Usually, if $\left[\frac{n}{2}\right]=k$ we rename the first $2k$ indeterminates
and express $\alpha$ as follows for $n=2k$ and  $n=2k+1$, respectively:
$$
\alpha'= dx -\sum_{1\leq i\leq k}(\xi_id\eta_i +\eta_id\xi_i) \text{  or  }
\alpha'= dx -\sum_{1\leq i\leq k}(\xi_id\eta_i +\eta_id\xi_i) -\zeta
d\zeta.
$$

On $(\Cee^*)^{1|n}$, there are 4 series of simple \lq\lq stringy" Lie
superalgebras of vector fields and 4 exceptional such superalgebras. The 13
of them
are distinguished: they admit nontrivial central extensions.

The \lq\lq main" 3 series are: $\fvect^L(1|n)=\fder\Cee[x^{-1}, x,\theta]$,
of all
vector fields, its subalgebra $\fsvect_{\lambda}^L(1|n)$ of vector fields that
preserve the volume form
$x^{\lambda}\vvol$, and $\fk^L(1|n)$ that preserves the Pfaff equation
$\alpha=0$. The superscript ${}^L$ indicates that we consider vector
fields with Laurent coefficients, not polynomial ones.

The Lie superalgebras of these 3 series are simple  with the exception of
$\fsvect_{\lambda}^L(1|1)$ for any $\lambda$, $\fsvect_{0}^L(1|n)$ for
$n>1$ and $\fk^L(1|4)$.

It so happens that $\fsvect_{0}^L(1|n)$ contains a simple
ideal of codimension $\varepsilon^n$, the quotient being spanned by
$\theta_1\dots\theta_n\partial_x$. Denote this ideal by
$\fsvect^{\circ L}(1|n)$; this is the fourth series.

The {\it twisted supercircle} of dimension $1|n$ is the supermanifold
that we denote $S^{1|n-1; M}$ is associated with the Whitney sum of the trivial
vector bundle of rank $n-1$ and the Moebius bundle. Since the Whitney sum
of the two
 Moebius bundles is isomorphic to the trivial rank 2 bundle, we will only
consider
either $S^{1|n}$ or $S^{1|n-1; M}$.

Let
$\theta^+_n=\sqrt{x}\theta_n$ be the corresponding to the Moebius bundle odd
coordinate on $\Cee S^{1|n-1; M}$, the complexification of $S^{1|n-1; M}$. Set
$$
\tilde\alpha=dx-\sum_{1\leq i\leq n-1}\theta_id\theta_i-x\theta_nd\theta_n;
$$
$$
\tilde\alpha'= dx -\sum_{1\leq i\leq k}(\xi_id\eta_i
+\eta_id\xi_i)-x\theta_nd\theta_n; \text{  or  }
\tilde\alpha'= dx -\sum_{1\leq i\leq k}(\xi_id\eta_i +\eta_id\xi_i) -\zeta
d\zeta-x\theta_nd\theta_n.
$$and define the fifth series  as the Lie superalgebra $\fk^M(n)$ that
preserves the Pfaff equation $\tilde\alpha=0$.

One exceptional superalgebra, $\fm^L(1)$, is the Lie subsuperalgebra in
$\fvect^L(1|1)$ that preserves the contact form
$$
\beta=d\tau+\pi dq-qd\pi
$$
corresponding to the \lq\lq odd mechanics" on $1|2$-dimensional 
supermanifold.  Though the following regradings demonstrate the 
isomorphism of this superalgebra with the nonexceptional ones, {\it 
considered as abstract} super algebras, they are distinct as filtered 
superalgebras and to various realizations of these Lie superalgebras different 
Schr\"odinger and KdV operators correspond.

Let $t, \xi$ be the indeterminates for $\fvect(1|1)$; let $t, \xi, 
\eta$ be same for $\fk (1|2)$ (in the realization that preserves the 
Pfaff eq.  $\alpha '=0$); and let $\tau, q, \xi$ be the indeterminates 
for $\fm(1)$.  Denote $\fvect (t, \xi)$ with the grading $\deg t=2$, 
$\deg\xi=1$ by$\fvect (t, \xi; 2, 1)$, etc.  Then the following 
exceptional nonstandard degrees indicated after a semicolon provide us 
with the isomorphisms:
$$
\begin{array}{cc}
\fvect (t,  \xi; 2, 1) \cong \fk (1|2);\; \; &\fk (t, \xi, \eta; 1,  2,  -1)
\cong
\fm (1);\\
\fvect (t,  \xi; 1, -1) \cong \fm (1);\; \; & \fm (\tau,  q,  \xi; 1,  2,
-1) \cong
\fk (1|2).
\end{array}
$$
 
Another exception is the Lie superalgebra $\fk^{L\circ}(1|4)$, the simple ideal of
codimension 1 in $\fk^L(1|4)$, the quotient being generated by
$\frac{\theta_1\theta_2\theta_3\theta_4}{x}$. The remaining exceptions are not
distinguished, so we ignore them in this paper.

\ssec{1.2.  The modules of tensor fields} To advance further, we have to
recall the
definition of the modules of tensor fields over the general vectoral Lie
superalgebra $\fvect(m|n)$ and its subalgebras, see [BL]. Let
$\fg=\fvect(m|n)$ (for any other $\Zee$-graded vectoral Lie superalgebra
the construction is
identical) and
$\fg_{\geq}=\mathop{\oplus}\limits_{i\geq 0}\fg_{i}$. Clearly,
$\fvect_0(m|n)\cong \fgl(m|n)$.
Let $V$ be the $\fgl(m|n)$-module with the {\it lowest} weight
$\lambda=\lwt(V)$. Make
$V$ into a $\fg_{\geq}$-module setting $\fg_{+}\cdot V=0$ for
$\fg_{+}=\mathop{\oplus}\limits_{i> 0}\fg_{i}$. Let us realize $\fg$ by
vector fields on the
$m|n$-dimensional linear supermanifold $\cC^{m|n}$ with coordinates $x=(u,
\xi)$. The
superspace $T(V)=\Hom_{U(\fg_{\geq})}(U(\fg), V)$ is isomorphic, due to the
Poincar\'e--Birkhoff--Witt theorem, to ${\Cee}[[x]]\otimes V$. Its elements
have a natural interpretation as formal {\it tensor fields of type} $V$ (or
$\lambda$). When $\lambda=(a, \dots , a)$ we will simply write $T(\vec a)$
instead of $T(\lambda)$. We usually consider irreducible $\fg_0$-modules.

{\it Examples}: $T(\vec 0)$ is the superspace of functions; $\Vol(m|n)=T(1,
\dots , 1;
-1, \dots , -1)$ (the semicolon separates the first $m$ coordinates of the
weight with respect to the matrix units $E_{ii}$ of $\fgl(m|n)$) is the
superspace of {\it densities} or {\it volume forms}. We denote the generator
of $\Vol(m|n)$ corresponding to the ordered set of coordinates $x$ by
$\vvol(x)$. The space
of $\lambda$-densities is $\Vol^{\lambda}(m|n)=T(\lambda, \dots , \lambda;
-\lambda, \dots , -\lambda)$. In particular, $\Vol^{\lambda}(m|0)=T(\vec
\lambda)$ but
$\Vol^{\lambda}(0|n)=T(\overrightarrow{-\lambda})$.

\ssec{1.3. Modules of tensor fields over stringy superalgebras}
Denote by $T^L(V)= \Cee[t^{-1}, t]\otimes V$ the
$\fvect(1|n)$-module that differs from $T(V)$ by allowing the Laurent
polynomials as
coefficients of its elements instead of just polynomials. Clearly,
$T^L(V)$ is a $\fvect^L(1|n)$-module. Define
the {\it twisted with weight $\mu$}\index{tensot fields, twisted} version of
$T^L(V)$ by setting:
$$
T^L_\mu (V)=\Cee[t^{-1}, t]t^\mu\otimes V.\eqno{(1.3.1)}
$$

$\bullet$ {\bf The \lq\lq simplest" modules --- the analogues of the
standard or identity representation of the matrix algebras}.  The simplest
modules over the Lie superalgebras of series $\fvect$ are,
clearly, the modules of $\lambda$-densities, $\Vol^{\lambda}$. These modules
are characterized by the fact that they are of rank 1 over $\cF$, the algebra of
functions. Over stringy superalgebras, we can also twist these modules and
consider
$\Vol^\lambda_\mu$. Observe that for $\mu\not\in\Zee$ this module has only
one submodule, the
image of the exterior differential $d$, see [BL], whereas for $\mu\in\Zee$
there is,
additionally, the kernel of the residue:
$$
\Res: \Vol^L \longrightarrow
\Cee \, , \;\;\; f\vvol_{t, \xi} \mapsto \text{ the coefficient of }\
\frac{\xi_{1} \ldots \xi_{n}}{t} \ \text{ in the expantion of }\
f.\eqno{(1.3.2)}
$$

$\bullet$ Over $\fsvect^L(1|n)$ all the spaces $\Vol ^\lambda$ are, clearly,
isomorphic, since their generator, $\vvol(t, \theta)$, is preserved. So all
rank 1 modules over
the module of functions are isomorphic to the module of twisted functions
$\cF_\mu$.

Over $\fsvect_\lambda^L(1|n)$, the simplest modules are generated by
$t^\lambda\vvol(t,
\theta)$. The submodules of the simplest modules over $\fsvect^L(1|n)$ and
$\fsvect_\lambda^L(1|n)$ are the same as those over $\fvect^L(1|n)$ but if
$\mu\in\Zee$ there is,
additionally, the trivial submodule generated by (the $\lambda$-th power
of) $\vvol(t, \theta)$
or $t^\lambda\vvol(t, \theta)$, respectively

$\bullet$ Over contact superalgebras $\fk(2n+1|m)$, it is more natural to
express
the simplest modules not in terms of $\lambda$-densities but via powers of
the form $\alpha$ which in what follows replaces $\alpha'$ for the $\fk^L$
series, or
$\tilde\alpha$ for the $\fk^M$ series, or
$\beta$ for $\fm^L(1)$:
$$
\cF_\lambda=\cases \cF\alpha^\lambda&\text{ for }n=m=0\\
\cF\alpha^{\lambda/2}&\text{otherwise }.\endcases\eqno{(1.3.3)}
$$
Observe that $\Vol ^\lambda\cong\cF_{\lambda(2n+2-m)}$ as
$\fk(2n+1|m)$-modules. In particular, the Lie superalgebra of series $\fk$
does not
distinguish between $\frac{\partial}{\partial t}$ and $\alpha^{-1}$: their
transformation rules are identical. Hence, 
$$
\fk(2n+1|m)\cong\cases \cF_{-1}&\text{ if }n=m=0\\
\cF_{-2}&\text{otherwise }.\endcases 
$$

$\bullet$ For $n=0, m=2$ (we take $\alpha =dt-\xi d\eta-\eta d\xi$) there
are other
rank 1 modules over $\cF$, the algebra of functions, namely:
$$
T(\lambda,
\nu)_\mu=\cF_{\lambda;\mu}\cdot\left(\frac{d\xi}{d\eta}\right)^{\nu/2}.\eqno
{(1.3.4)}
$$

$\bullet$ Over $\fk^M$, we should replace the form $\alpha$ with
$\tilde \alpha$ and the definition of the $\fk^L(1|m)$-modules
$\cF_{\lambda; \mu}$ should be
replaced with
$$
\cF^M_{\lambda; \mu}=\cases \cF_{\lambda; \mu}(\tilde
\alpha)^\lambda&\text{ for }m=1\\
\cF_{\lambda; \mu}(\tilde \alpha)^{\lambda/2}&\text{ for
}m>1.\endcases\eqno{(1.3.5)}
$$

$\bullet$ For $m=3$ and $\alpha =dt-\xi d\eta-\eta d\xi-t\theta d\theta$
there are
other rank 1 modules over the algebra of functions $\cF$, namely:
$$
T^M(\lambda, \nu)_\mu=\cF^M_{\lambda,
\mu}\cdot\left(\frac{d\xi}{d\eta}\right)^{\nu/2}.\eqno{(1.3.6)}
$$

\begin{rem*}{Examples} 1) The $\fk(2n+1|m)$-module of volume
forms is $\cF _{2n+2-m}$. In particular, $\fk(1|2) \subset
\fsvect(1|2)$.

2) As $\fk^L(1|m)$-module, $\fk^L(1|m)$ is isomorphic to $\cF _{-1}$ for
$m=0$ and $\cF _{-2}$ otherwise. As $\fk^M(1|m)$-module, $\fk^M(1|m)$ is
isomorphic to
$\cF _{-1}$ for $m=1$ and $\cF _{-2}$ otherwise. In particular,
$\fk^L(1|4) \simeq \Vol$ and $\fk^M(1|5) \simeq \Pi(\Vol)$.
\end{rem*}

\ssec{1.4. Convenient formulas} The four series of classical stringy
superalgebras
are $\fvect^L(1|n)$,
$\fsvect^L_\lambda(1|n)$, $\fk^{L}(1|n)$ and  $\fk^{M}(1|n)$.
$$
D=f\partial_t+\sum f_i\partial_i\in \fsvect^{L}_{\lambda}(1|n)\quad\text{if
and only
if }\quad \lambda f=-t\Div D.\eqno{(1.4.1)}
$$

A laconic way to describe
$\fk$, $\fm$ and their subalgebras is via {\it generating functions}.

$\bullet$ Odd form $\alpha=\alpha_1$. For $f\in\Cee [t, \theta]$ set\index{$K_f$
contact vector field} \index{$H_f$ Hamiltonian vector field}:
$$
K_f=\triangle(f)\pder{t}-H_f +
\pderf{f}{t} E,
$$
where
$E=\sum\limits_i \theta_i
\pder{\theta_{i}}$, $\triangle (f)=2f-E(f)$, and $H_f$ is the
hamiltonian field with Hamiltonian $f$ that preserves $d\alpha_1$:
$$
H_f=-(-1)^{p(f)}\left(\sum\limits_{j\leq m}\pderf{
f}{\theta_j} \pder{\theta_j}\right ) , \; \; f\in \Cee [\theta].
$$

The choice of the form $\alpha'$ instead of $\alpha$ only affects the
form of $H_f$ that we give for $m=2k+1$:
$$
H_f=-(-1)^{p(f)}\sum\limits_{j\leq
k}(\pderf{f}{\xi_j} \pder{\eta_j}+
\pderf{f}{\eta_j} \pder{\xi_j}+
\pderf{f}{\theta} \pder{\theta}), \;
\; f\in \Cee [\xi, \eta, \theta].
$$

$\bullet$ Even form $\beta=\alpha_0$. For $f\in\Cee [q, \xi, \tau]$ set:
$$
M_f=\triangle(f)\pder{\tau}- Le_f
-(-1)^{p(f)} \pderf{f}{\tau} E,
$$
where $E=\sum\limits_iy_i
\pder{y_i}$ (here the $y$ are all the coordinates except
$\tau$) is the Euler operator, $\triangle(f)=2f-E(f)$, and
$$
Le_f=\sum\limits_{i\leq n}( \pderf{f}{q_i}\
\pder{\xi_i}+(-1)^{p(f)} \pderf{f}{\xi_i}\
\pder{q_i}), \; f\in \Cee [q, \xi].
$$
Since
$$
L_{K_f}(\alpha_1)=K_1(f)\alpha_1, \quad\quad
L_{M_f}(\alpha_0)=-(-1)^{p(f)}M_1(f)\alpha_0,\eqno{(1.4.2)}
$$
it follows that $K_f\in \fk (2n+1|m)$ and $M_f\in \fm (n)$. Observe that
$$
p(Le_f)=p(M_f)=p(f)+\od.
$$

$\bullet$ To the (super)commutators $[K_f, K_g]$ or $[M_f, M_g]$ there
correspond {\it contact brackets}\index{Poisson bracket}\index{contact bracket}
of the generating functions:
$$
[K_f, K_g]=K_{\{f, g\}_{k.b.}};\quad\quad [M_f, M_g]=M_{\{f, g\}_{m.b.}}
$$
The explicit formulas for the contact brackets are as follows. Let us first
define the brackets on functions that do not depend on $t$ (resp. $\tau$).
The {\it Poisson bracket} $\{\cdot , \cdot\}_{P.b.}$ is given by the formula
$$
\begin{array}{l}
{}\{f, g\}_{P.b.}=-(-1)^{p(f)}\sum\limits_{j\leq m}\
\pderf{f}{\theta_j}\ \pderf{g}{\theta_j}\; \text{or}\\
{}\{f, g\}_{P.b.}=-(-1)^{p(f)}[\sum\limits_{j\leq m}(
\pderf{f}{\xi_j}\ \pderf{
g}{\eta_j}+\pderf{f}{\eta_j}\ \pderf{
g}{\xi_j})+\pderf{f}{\theta}\ \pderf{
g}{\theta}].
\end{array}\eqno{(1.4.3)}
$$
The {\it Buttin bracket} $\{\cdot ,
\cdot\}_{B.b.}$ \index{Buttin bracket $=$ Schouten bracket} is given by the
formula
$$
\{ f, g\}_{B.b.}=\sum\limits_{i\leq n}\ (\pderf{f}{q_i}\
\pderf{g}{\xi_i}+(-1)^{p(f)}\ \pderf{f}{\xi_i}\
\pderf{g}{q_i}).\eqno{(1.4.4)}
$$
In terms of the Poisson and Buttin brackets, respectively, the contact
brackets take the form
$$
\{ f, g\}_{k.b.}=\triangle (f)\pderf{g}{t}-\pderf{f}
{t}\triangle (g)-\{ f, g\}_{P.b.}\eqno{(1.4.5)}
$$
and, respectively,
$$
\{ f, g\}_{m.b.}=\triangle (f)\pderf{g}{\tau}+(-1)^{p(f)}
\pderf{f}{\tau}\triangle (g)-\{ f, g\}_{B.b.}.\eqno{(1.4.6)}
$$

It is not difficult to prove the following isomorphisms (as superspaces):
$$
\fk (2n+1|m)\cong\Span(K_f: f\in \Cee[t, p, q, \xi]);\quad
\fm (n)\cong\Span(M_f: f\in \Cee [\tau, q, \xi]).
$$

It is not difficult to verify that $\fk^{M}(1|n) =\Span(\tilde K_{f}: f\in
R^{L}(1|n))$,
where the M\"obius contact field is given by the formula
$$
\tilde K_{f} =\bigtriangleup (f) {\cal D}+ {\cal D}(f)E
+\tilde H_{f},\eqno{(1.4.7)}
$$
in which $\bigtriangleup = 2 - E$, but where
$$
E= \sum\limits_{i\leq n}\theta_{i}
\pder{\theta_{i}}+\frac 12\theta\pder{\theta}\quad\text{and}\quad{\cal D}
=\displaystyle
\pder{ t} -\displaystyle\frac{\theta}{2t}\displaystyle\pder{
\theta}=\frac12\tilde K_1
$$
and where
$$
\tilde H_{f}\; =\; (-1)^{p(f)}(\sum \;
\pderf{f}{\theta_{i}}
\pder{\theta_{i}}
+\; \frac{1}{t} \pderf{f}{\theta}\pder{\theta}) 
$$
in the realization
with form $\tilde\alpha$; in the realization
with form $\tilde \alpha'$ for $n=2k$ and $n=2k+1$ we have, 
respectively:
$$
\begin{matrix}
\tilde H_{f}\; =\; (-1)^{p(f)}(\sum \;
(\pderf{f}{\xi_{i}}
\pder{\eta_{i}}+\pderf{f}{\eta_{i}}
\pder{\xi_{i}})
+\; \frac{1}{t} \pderf{f}{\theta}\pder{\theta}) ,\\
\tilde H_{f}\; =\; (-1)^{p(f)}(\sum \;
(\pderf{f}{\xi_{i}}
\pder{\eta_{i}}+\pderf{f}{\eta_{i}}
\pder{\xi_{i}})+\pderf{f}{\zeta}
\pder{\zeta}
+\; \frac{1}{t} \pderf{f}{\theta}\pder{\theta}) .
\end{matrix}
$$
The corresponding contact bracket of generating functions will be called the
{\it Ramond bracket}; its explicit form is
$$
\{f, g\}_{R.b.}=\triangle (f)\cD(g) -
\cD(f)\triangle (g)-\{f, g\}_{MP.b.},\eqno{(1.4.8)}
$$
where the {\it M\"obius-Poisson bracket}\index{ M\"obius-Poisson
bracket}\index{bracket M\"obius-Poisson} $\{\cdot , \cdot\}_{MP.b}$ is
$$
\{f, g\}_{MP.b}=(-1)^{p(f)}\left (\sum \;
\pderf{f}{\theta_{i}}
\pderf{g}{\theta_{i}}
+\; \frac{1}{t} \pderf{f}{\theta}\pderf{g}{\theta}\right) \quad\text{in the
realization
with form $\tilde\alpha$}.\eqno{(1.4.9)}
$$

Observe that
$$
L_{K_{f}}(\alpha_1)=K_1(f)\alpha_1, \quad
L_{\tilde K_{f}}(\tilde \alpha)=\tilde K_1(f)\tilde\alpha.\eqno{(1.4.10)}
$$

\ssec{1.5. Distinguished stringy superalgebras}

\begin{Theorem} The only nontrivial central extensions of the simple stringy Lie
superalgebras are those given in the following table. 
\end{Theorem}
The operator $\nabla$ introduced in the second column of the table by 
the formula $c: D_1, D_2\mapsto \Res (D_1, \nabla(D_2))$ for an 
appropriate pairing $(\cdot, \cdot)$ will 
be referred to as the {\it cocycle operator}.

Let in this sebsection and in sec.  2.1 $K_f$ be the common notation 
of both $K_f$ and $\tilde K_f$, depending on whether we consider 
$\fk^{L}$ or $\fk^{M}$, respectively.  Let further $\cK = 
(2\theta\pder{\theta} - 1)\pder{x^2}^2$ and let $c: K_{f}, K_{g}\mapsto 
\Res (K_{f}, \nabla(K_{g}))$ or $c: M_{f}, M_{g}\mapsto 
\Res (M_{f}, \nabla(M_{g}))$ be the cocycle that determines the 
nontrivial central extention.
$$
\renewcommand{\arraystretch}{1.4}
\begin{tabular}{|c|c|c|}
\hline
algebra & the cocycle $c$ & The extended algebra \\
\hline
$\fk^{L}(1|0)$ & $\Res fK_{1}^3(g)$ & Virasoro or $\;
\fvir$ \\
\hline
$\left. \begin{matrix}\fk^{L}(1|1) \\
\fk^{M}(1|1) \end{matrix}\right\}$
 & $\Res fK_{\theta }K_{1}^2(g)$
 & $\begin{matrix} \text{Neveu-Schwarz or} \; \fns \\
    \text{Ramond or}\; \fr \end{matrix}$ \\
\hline
$ \left.\begin{matrix}\fk^{L}(1|2)\\
\fk^{M}(1|2) \end{matrix}\right\}$
 & $(-1)^{p(f)}\Res fK_{\theta_1 }K_{\theta_2}K_{1}(g)$
 & $\begin{matrix}\text{2-Neveu-Schwarz or}\;\fns(2) \\ 
\text{2-Ramond or}\;\fr(2)\end{matrix}$ \\
\hline
$\fm^{L}(1)$
 & $M_{f}, M_{g}\mapsto \Res f(M_{\xi})^3(g)$
 & $\widehat{\fm^{L}(1)}$ \\
\hline
$\left.\begin{matrix}\fk  ^{L}(1|3) \\ 
\fk^{M}(1|3)\end{matrix} \right\}$
 & $\Res fK_{\xi }K_{\theta }K_{\eta }(g)$
 & $\begin{matrix}\text{3-Neveu-Schwarz or}\; \fns(3) \\ 
\text{3-Ramond or}\; \fr (3)\end{matrix}$ \\
\hline
$ \left.\begin{matrix}\fk^{L\circ}(4) \\ 
\fk^{M}(1|4) \end{matrix}\right\}$ & $(1)\; (-1)^{p(f)}\Res 
fK_{\theta_{1}}K_{\theta_{2}}K_{\theta_{3}}K_{\theta 
_{4}}K_{1}^{-1}(g) $
 & $\begin{matrix}\begin{matrix}\text{4-Neveu-Schwarz}=\fns(4)\\  
\text{$4$-Ramond}=\fns(4)\end{matrix}\\  
    \end{matrix}$ \\ 
\hline 
$ \fk^{L\circ}(4) $ & $\begin{matrix}(2) & \Res f(tK_{t^{-1}}(g)) \\ 
   (3) & \Res fK_{1}(g)\end{matrix}$
 & $\begin{matrix}  \text{$4'$-Neveu-Schwarz}=\fns(4') \\
  \text{$4^0$-Neveu-Schwarz}=\fns(4^0) 
   \end{matrix}$ \\ 
\hline \end{tabular}   
$$
Observe that $K_{1}^{-1}$ is only defined on $\fk^{L\circ}(4)$ but 
not on $\fk^{L}(4)$; though functions $(2)\;  \Res f(tK_{t^{-1}}(g))$ 
and $(3)\; \Res fK_{1}(g)$ are defined on $\fk^{L}(4)$ and 
$\fk^{M}(4)$ they are not even cocycles there.

$$ 
\renewcommand{\arraystretch}{1.3}
\begin{tabular}{|c|c|c|}
\hline
& the restriction of the cocycle (1) on $\fk^{L\circ}(4)$ : & \\
$\fvect ^{L}(1|2)$ & $D_1=f\pder{t}+g_{1}\pder{\xi
_{1}}+g_{2} \pder{ \xi _{2}}, \;
D_2=\tilde{f}\pder{ t}+\tilde{g}_{1}\pder{ \xi
_{1}}+\tilde{g}_{2}\pder{
\xi_{2}}$& $\widehat{\fvect}^{L}(1|2)$ \\  
&$\mapsto\;\Res(g_{1}\tilde{g}_{2}'-g_{2}\tilde{g}_{1}'(-1)^{p(D_{1})p(D_{2})})$&\\
\hline
$\fsvect ^{L}_{\lambda }(1| 2)$& the restriction of the above  
& $\widehat{\fsvect}^{L}_{\lambda}(1|2)$ \\
\hline
$\fvect ^{L}(1|1)$ & $D_1=f\pder{t}+g\pder{ \xi}, \;
D_2=\tilde{f}\pder{ t}+\tilde{g}\pder{ \xi}\mapsto\;$& \\
&$\Res (f\cK(\tilde g) + (-1)^{p(D_1)p(D_2)}g\cK(\tilde f)
+$&$\widehat{\fvect}^{L}(1|1)$\\
&$2(-1)^{p(D_1)p(D_2)+p(D_2)}g\pder{\theta}\pder{\theta}(\tilde
g)$&\\
\hline   \end{tabular}   
$$

Observe that the restriction of the only cocycle on
$\fvect^{L}(1| 2)$ to its subalgebra $\Span(f(t)\pder{t})\cong\fwitt$ is trivial
while the the restriction of the only cocycle on $\fsvect ^{L}_{\lambda }(1|2)$ to its unique
subalgebra $\fwitt$ is nontrivial. The riddle is solved by a closer study of the embedding
$\fvect(1|m)\tto\fk(1|2m)$: it involves differentiations.

Explicitely, the embedding $i:\fvect^L(1|n)\tto\fk^L(1|2n)$ is given by
the following formula in which
$\Phi=\sum\limits_{i\leq n}\xi_i\eta_i$ (if we are not keen to preserve the
isomorphism but are only interested in a subalgebra of
$\fk^L(1|2n)$ isomorphic to
$\fvect^L(1|n)$, then the coefficients $\frac{(-1)^{p(f)}}{2^m}$ can be dropped
to simplify life):
$$
\renewcommand{\arraystretch}{1.3}
\begin{tabular}{|c|c|}
\hline
$D\in\fvect^L(1|n)$ &the generator of $i(D)$\cr
\hline
$f(\xi)t^m\partial_t$&$(-1)^{p(f)}\frac{1}{2^m}f(\xi)(t-\Phi)^m$\cr
$f(\xi)t^m\partial_i$&$
(-1)^{p(f)}\frac{1}{2^m}f(\xi)\eta_i(t-\Phi)^{m}$\cr
\hline
\end{tabular}\eqno{(1.5.1)}
$$

Clearly,
$\fsvect^L_\lambda(1|n)$ is the subsuperspace of
$\fvect^L(1|n)$ spanned by the expressions
$$
f(\xi)(t-\Phi)^m+\sum\limits_i
f_i(\xi)\eta_i(t-\Phi)^{m-1}\; \text{such that}\;
(\lambda +n)f(\xi)=-\sum\limits_i (-1)^{p(f_{i})}\pderf{f_i}{\xi_i}.
\eqno{(1.5.2)}
$$

The nonzero values of the cocycle $c$ on $\fvect^{L}(1| 2)$ in monomial
basis are:
$$
\renewcommand{\arraystretch}{1.7}
\begin{matrix}
\displaystyle c(t^k\theta_1\pder{\theta_1},
t^l\theta_2\pder{\theta_2})=k\delta_{k, -l},&
\displaystyle c(t^k\theta_1\pder{\theta_2}, t^l\theta_2\pder{\theta_1})=-k\delta_{k, -l},\\
\displaystyle c(t^k\theta_1\theta_2\pder{\theta_1},
t^l\pder{\theta_2})=k\delta_{k, -l},&
\displaystyle c(t^k\theta_1\theta_2\pder{\theta_2},
t^l\pder{\theta_1})=k\delta_{k, -l}.
\end{matrix} \eqno{(1.5.3)}
$$

In $\fsvect ^{L}_{\lambda}(1|2)$, set:
$$
\renewcommand{\arraystretch}{1.6}
\begin{matrix}
\displaystyle L_m=t^m\left
(t\pder{t}+\frac{\lambda+m+1}{2}(\theta_1\pder{\theta_1}+
\theta_2\pder{\theta_2})\right),\\
\displaystyle S_m^j=t^m\theta_j\left
(t\pder{t}+(\lambda+m+1)(\theta_1\pder{\theta_1}+
\theta_2\pder{\theta_2})\right).
\end{matrix} \eqno{(1.5.4)}
$$

The nonzero values of the cocycles on $\fsvect ^{L}_{\lambda }(1|2)$ are
$$
\renewcommand{\arraystretch}{1.6}
\begin{matrix}
\displaystyle c(L_m, L_n)=\frac 12m(m^2-(\lambda+1)^2)\delta_{m, -n},\\
\displaystyle c(t^k\pder{\theta_i}, S_m^j)=-m(m-(\lambda+1))\delta_{m,
-n}\delta_{i, j},\\
\displaystyle c(t^m(\theta_1\pder{\theta_1}-\theta_2\pder{\theta_2}),
t^n(\theta_1\pder{\theta_1}-\theta_2\pder{\theta_2}))=m\delta_{m, 
-n},\\
\displaystyle c(t^m\theta_1\pder{\theta_2},
t^n\theta_2\pder{\theta_1})=m\delta_{m, -n}.
\end{matrix} \eqno{(1.5.5)}
$$

\section*{\protect \S 2. Superized KdV and Schr\"odinger operators}

\ssec{2.1.  The KdV operators for the distinguished contact 
superalgebras} Let in this subsection $K_f$ be the common term for 
both $K_f$ and $\tilde K_f$, as Table 1.5, for $\fg=\fk^{L}$ or 
$\fk^{L\circ}$ or $\fk^{M}$.  The equation for $K_f\in\fst_{(F, 1)}$, 
where $(F, 1)\in\hat \fg^*$, is of the form $KdV(f)=0$, where the 
operators $KdV$ are listed in the following table with the cocycle 
operators being the symmetrizations of the operators $\nabla$ from 
sec.  1.5.
$$
\renewcommand{\arraystretch}{1.4}
\begin{tabular}{|c|c|}
\hline
$n$&$KdV=$ the cocycle operator$\oplus$ \lq\lq the standard part"\\
\hline
$0$&$\frac{d^3}{x^3}\oplus F\frac{d}{x}+
\frac{d}{dx}F$\\
\hline
$1$&$
K_{\theta}(K_1)^2\oplus 2(F\partial_x+\partial_xF)+(-1)^{p(F)}K_{\theta}
FK_{\theta}$\\
\hline
$2$&$(K_{\xi}K_{\eta}-
K_{\eta}K_{\xi})K_1\oplus 2(F\partial_x+\partial_xF)+(-1)^{p(F)}(K_{\xi}FK_{\eta}-K_{\eta}FK_{\xi})$\\
\hline
$3$&
$(K_{\xi}K_{\eta}-K_{\eta}K_{\xi})K_{\theta}\oplus
$\\
&$2F\partial_x+2\partial_xF+(-1)^{p(F)}(K_{\xi}FK_{\eta}-K_{\eta}FK_{\xi}+K_{\theta}FK_{\theta})$\\
\hline
$4_1$&$(K_{\xi_1}K_{\eta_1}-K_{\eta_1}K_{\xi_1})(K_{\xi_2}K_{\eta_2}-
K_{\eta_2}K_{\xi_2})\int_x\oplus$\\
&$2(F\partial_x+\partial_xF)+(-1)^{p(F)}\sum_{i=1,
2}(K_{\xi_{i}}FK_{\eta_{i}}-K_{\eta_{i}}FK_{\xi_{i}})$\\
\hline
$4_2$&$xK_{x^{-1}}\oplus$ the standard part from $4_1$\\
\hline
$4_3$&$K_1\oplus$ the standard part from $4_1$\\
\hline
\end{tabular}
$$

The KdV operator can, therefore, be always represented in the 
following form with \lq\lq the standard part" explicified:
$$
\text{the cocycle operator}+
2(F\partial_x+\partial_xF)+(-1)^{p(F)}\sum(K_{\xi_{i}}FK_{\eta_{i}}-K_{\eta_
{i}}FK_{\xi_{i}})\eqno{(\text{for}\;
\fk^L(1|n))}
$$
$$
\begin{matrix}
\text{the
cocycle operator}+2(F\partial_x+\partial_xF)- F\frac{\theta}{x}\partial_\theta-
\frac{\theta}{x}\partial_\theta F+\frac{F(1-E)}{x}+\\
(-1)^{p(F)}[\sum_{i\leq
n-1}K_{\theta_{i}}FK_{\theta_{i}}+tK_{\theta}FK_{\theta}]+
\frac{1}{2x}[\theta\partial_\theta F\sum_{i\leq
n-1}\theta_{i}\partial_{\theta_{i}}-
\sum_{i\leq n-1}\theta_{i}\partial_{\theta_{i}}
\theta\partial_{\theta}].
\end{matrix}\eqno{(\text{for}\; \fk^M(1|n))}
$$
So the KdV operators corresponding to the supercircles associated with 
the cylinder and the M\"obius
band are absolutely different. To establish that similar is the situation
with the
Schr\"oedinger operators, let us compair $\fg$ with $\fg^*$ for the
cylinder and the M\"obius
band:
$$
\renewcommand{\arraystretch}{1.4}
\begin{tabular}{|c|c|c|c|c|c|c|c|c|c|}
\hline
$\fg=\fk^L(1|n)$&0&1&2&3&4&5&6&7\\
\hline
$\fg$&$\cF_{-1}$&$\cF_{-2}$&$\cF_{-2}$&$\cF_{-2}$&
$\cF_{-2}$&$\cF_{-2}$&$\cF_{-2}$&$\cF_{-2}$\\
\hline
$\Vol$&$\cF_{1}$&$\Pi(\cF_{0})$&$\cF_{-1}$&$\Pi(\cF_{-2})$&
$\cF_{-3}$&$\Pi(\cF_{-4})$&$\cF_{-5}$&$\Pi(\cF_{-6})$\\
\hline
$\fg^*$&$\cF_{2}$&$\Pi(\cF_{3})$&$\cF_{2}$&$\Pi(\cF_{1})$&
$\cF_{0}$&$\Pi(\cF_{-1})$&$\cF_{-2}$&$\Pi(\cF_{-3})$\\
\hline
\end{tabular}
$$

$$
\renewcommand{\arraystretch}{1.4}
\begin{tabular}{|c|c|c|c|c|c|c|c|c|c|}
\hline
$\fg=\fk^M(1|n)$&1&2&3&4&5&6&7&$n$\\
\hline
$\fg$&$\cF_{-1}$&$\cF_{-2}$&$\cF_{-2}$&$\cF_{-2}$&
$\cF_{-2}$&$\cF_{-2}$&$\cF_{-2}$&$\cF_{-2}$\\
\hline
$\Vol$&$\Pi(\cF_{1})$&$\cF_{1}$&$\Pi(\cF_{0})$&
$\cF_{-1}$&$\Pi(\cF_{-2})$&$\cF_{-3}$&$\Pi(\cF_{-4})$&$\Pi^n(\cF_{3-n})$\\
\hline
$\fg^*$&$\Pi(\cF_{2})$&$\cF_{3}$&$\Pi(\cF_{2})$&
$\cF_{1}$&$\Pi(\cF_{0})$&$\cF_{-1}$&$\Pi(\cF_{-2})$&$\Pi^n(\cF_{5-n})$\\
\hline
\end{tabular}
$$

The comparison of $\fg$ with $\fg^*$ shows that there is a nondegenerate
bilinear form
on
$\fg=\fk^L(1|6)$ and
$\fg=\fk^M(1|7)$, even and odd, respectively. These forms are supersymmetric
and given by the formula
$$
(K_f, K_g)=\Res~fg.
$$

\ssec{2.2. Examples} $1_{\fns(1)}$) is considered in sec. 0.2.

$1_{\fr(1)}$) $\hat\fg=\fr(1)$ Let us calculate the stabilizer of an
element of $\fr(1)^*$. In doing so we will use the $C$-points of all objects
encountered. Observe again that since the integral (residue) pairs $1$ with
$\frac{\theta}{t}$, this pairing is odd, and, therefore,
$\fr(1)^*=\Pi(\cF _2)$.

In components the equation (for $n=1$) takes the form:
$$
\left[\begin{pmatrix}
L_3&0\\
0&L_2\end{pmatrix}+ \begin{pmatrix}
0&2F_0\frac{d}{dx}+\frac{d}{dx}F_0-\frac{F_0}{x}\\
F_0\frac{d}{dx}+2\frac{d}{dx}F_0+\frac{F_0}{x}&0\end{pmatrix}\right]\begin{pmatrix} 
f_0\\
f_1\end{pmatrix}=0,
$$
where $L_3=\frac{d^3}{dx^3}+2\frac{d}{dx}F_1+ 2F_1\frac{d}{dx}$,
$L_2=\frac{1}{x}(\frac{\partial}{\partial x}-\frac{1}{2x})^2-\frac{F_1}{x}$.

$2_{\fns(2)}$) In components the analog of KdV corresponding to
$\fns(2)$ is
$$
\begin{matrix}\left[ \begin{pmatrix}
L_3&2F_0\frac{d}{dx}&0&0\\
2F_0\frac{d}{dx}&\frac{d}{dx}&0&0\\
0&0&0&L_2\\
0&0&L_2&0\end{pmatrix}+ \right.\\
\left.\begin{pmatrix}
0&0&-(2F_2\frac{d}{dx}+\frac{d}{dx}F_2)&2F_1\frac{d}{dx}+\frac{d}{dx}F_1\\
0&0&F_2&-F_1\\
-(F_2\frac{d}{dx}+2\frac{d}{dx}F_2)&-F_2&0&0\\
F_1\frac{d}{dx}+2\frac{d}{dx}F_1&F_1&0&0\end{pmatrix} \right]\begin{pmatrix}
f_0\\
f_{12}\\
f_1\\
f_2\end{pmatrix}=0,
\end{matrix}
$$
where $f=f_0+f_1\xi+f_2\eta+f_{12}\xi \eta$, $F=F_0+F_1\xi+F_2\eta+F_{12}\xi
\eta$ and
$$
L_3=\frac{d^3}{dx^3}+2F_{12}\frac{d}{x}+2\frac{d}{dx}F_{12};\quad
L_2=\frac{d^2}{dx^2}+F_{12}+F_0\frac{d}{dx}+\frac{d}{dx}F_{0}.
$$

\ssec{2.3. The list of Sch\"rodinger operators}
In the following table we use an abbreviation: $\Delta=
\tilde K_\theta (\tilde K_1)^{-1}-(\tilde K_1)^{-1}\tilde K_\theta$. The
Sch\"rodinger
operator is the sum of the operator given in the tables with a potential
$F$, where
$F\in\cF$ or
$F\in\Pi(\cF)$: the parity of the potential should be equal to that of the
operator.

$$
\renewcommand{\arraystretch}{1.4}
\begin{tabular}{|c||c|c|c|c|}
\hline
$n$& $0$&$1$&$2$&$3$\\ \hline
$\fk^L(1|n)$&$K_1^2$&$K_\theta K_1$&$K_\xi K_\eta-K_\eta K_\xi$&$(K_\xi K_\theta
K_\eta-K_\eta K_\theta K_\xi)(K_1)^{-1}$\\
\hline
$\fk^M(1|n)$&$-$&$\Delta \tilde K_1^2$&$\Delta \tilde K_{\theta_{1}} \tilde
K_1$&$\Delta (\tilde K_\xi \tilde K_\eta-\tilde K_\eta \tilde K_\xi)$\\
\hline
\end{tabular}
$$

$$
\renewcommand{\arraystretch}{1.4}
\begin{tabular}{|c||c|}
\hline
$\fk^L(1|4)$&(1)\quad $(K_{\xi_{1}}
K_{\eta_{1}}-K_{\eta_{1}}K_{\xi_{1}})(K_{\xi_{2}}
K_{\eta_{2}}-K_{\eta_{2}}K_{\xi_{2}})(K_1)^{-2}$\\
&(2)\quad $xK_{x^{-1}} (K_1)^{-1}-(K_1)^{-1}xK_{x^{-1}}$\\
&(3)\quad ??\\ \hline
$\fk^M(1|4)$&(1)\quad $\Delta \tilde K_{\theta_{1}}(\tilde K_\xi \tilde
K_\eta-\tilde K_\eta \tilde K_\xi)(\tilde K_1)^{-1}$\\
\hline
\end{tabular}
$$
So far we failed to write an explicit expression for the third Sch\"rodinger
operators.

For the Lie superalgebra $\fvect^L(1|1)$ the Sch\"rodinger
operator is the same operator as for $\fk^L(1|2)$ but rewritten in the 
form of a matrix and with $\eta$ replaced with $\partial_{\xi}$. We 
leave as an exercise to the reader the pleasure to write this matrix 
explicitely as well as to reexpress it in terms of the fields $M_{f}$ 
for $\fm^{L}(1)$. 

For $\fvect^L(1|2)$ the Sch\"rodinger
operators are obtained from the Sch\"rodinger operator
$$
\begin{pmatrix}
\xi_1\xi_2\partial_{\xi_{2}}\partial^2_x&
\xi_2\partial_{\xi_{1}}-\partial_{\xi_{2}}\partial_x&
\xi_1\partial_{\xi_{1}}\partial_{\xi_{2}}\partial_x&
\partial_{\xi_{1}}\partial_{\xi_{2}}\\
-\xi_1\xi_2\partial_{\xi_{2}}\partial^3_{x}&
-\xi_1\xi_2\partial_{\xi_{1}}\partial_{\xi_{2}}\partial^2_x&
\xi_1\partial_{\xi_{2}}\partial^2_x&
-\xi_1\partial_{\xi_{1}}\partial_{\xi_{2}}\partial_x\\
\xi_1\xi_2\partial_{\xi_{1}}\partial^3_x&
\xi_1\partial_{\xi_{1}}\partial^2_x&
\xi_2\partial_{\xi_{2}}\partial^2_x&
\xi_2\partial_{\xi_{1}}\partial_{\xi_{2}}\partial_x\\
-\xi_1\xi_2\partial_{\xi_{1}}\partial^4_x&
-\xi_1\xi_2\partial_{\xi_{1}}\partial^3_x&
-\xi_1\xi_2\partial_{\xi_{2}}\partial^3_x&
-\xi_1\xi_2\partial_{\xi_{1}}\partial_{\xi_{2}}\partial^2_x\\
\end{pmatrix}
\partial^{-2}_x\eqno{(*)}
$$
for $\fk^L(1|4)$ after restriction to the subalgebra $\fvect^L(1|2)$, 
see (1.2).  In the dual space this leads to the quotient space: we 
should disregard the first row and the last column of $(*)$.

\ssec{2.4. The Schr\"odinger operators as selfadjoint differential 
operators}{}

\par 

{}$\bullet$ For the Neveu--Schwarz superalgebras we have the exact
sequences
$$
0\longrightarrow \fz\longrightarrow \fns (n)\longrightarrow\cF _2
\longrightarrow  0.\eqno{(2.4.1)}
$$
Here $\fz=\Span(z)$ if $n\neq 4$ and $\fz=\Span(z, z_1, z_2)$ if $n=4$.

Using the identification $\Vol\cong \Pi^n(\cF _{2-n})$ we
dualize the above exact sequence and get:
$$
\begin{matrix}
0\longrightarrow \Pi^n(\cF _{4-n})\longrightarrow \fns^*
(n)\longrightarrow \fz^*\longrightarrow  0\quad\text{for}\; n= 1, 2, 3,\\
0\longrightarrow \cF _{0}/\Cee\longrightarrow \fns^*
(4)\longrightarrow \fz^*\longrightarrow  0,\\
0\longrightarrow \cF _{0}/\Cee\longrightarrow \fns^*
(4')\longrightarrow \fz^*\longrightarrow  0,\\
0\longrightarrow \cF _{0}/\Cee\longrightarrow \fns^*
(4^0)\longrightarrow \fz^*\longrightarrow  0.\\
\end{matrix}\eqno{(2.4.2)}
$$
For the cocycles (2) and (3) from Table 1.5 we can as well consider not the
$\fns^*(4')$ and $\fns^* (4^0)$ but the algebras of their derivations; than
we do not have to factorize modulo constants which simplifies life.

$\bullet$ For the Ramond superalgebras we similarly have the exact
sequences
$$
\begin{matrix}
0\longrightarrow \fz\longrightarrow \fr (1)\longrightarrow
\cF _{-1}\longrightarrow  0&\\
0\longrightarrow \fz\longrightarrow \fr (n)\longrightarrow
\cF _{-2}\longrightarrow  0&\text{for $n>1$}.
\end{matrix}\eqno{(2.4.3)}
$$
Here $\fz=\Span(z)$.

Using the identification $\Vol\cong \begin{cases}\Pi(\cF _{1})&\text{for}\;
n=1\\
\Pi^n(\cF _{3-n})&\text{for}\;
n>1\end{cases}$ we dualize the above exact sequence and get:
$$
\begin{matrix}
0\longrightarrow \Pi(\cF _{2})\longrightarrow \fr^*
(1)\longrightarrow \fz^*\longrightarrow  0\\
0\longrightarrow \Pi^n(\cF _{5-n})\longrightarrow \fr^*
(n)\longrightarrow \fz^*\longrightarrow  0.
\end{matrix}\eqno{(2.4.4)}
$$

Let us realize the elements of $\fns^*(n)$ and $\fr^*(n)$ by selfadjoint
(pseudo)differential operators $\hat F: \cF _{\lambda}\longrightarrow
\Pi^n(\cF_{\mu})$. We have already done this for $\fvir$ in Introduction.
The order of
$\hat F$ is equal to $4-n$ for $\fns^*(n)$; it isequal to $5-n$ for
$\fr^*(n)$ if $n>1$ and $2$ for $\fr^*(1)$.

Now, let us solve the systems of two equations, of which the first equation
counts
the order of $\Sch$ and the second one is the dualization condition:
$$
\begin{matrix}
\mu=2+(2-n)+\lambda, &\mu+\lambda=2-n&\text{for  }\fns(n)\\
\mu=1+1+\lambda, &\mu+\lambda=1&\text{for  }\fr(1)\\
\mu=2+(3-n)+\lambda, &\mu+\lambda=3-n&\text{for  }\fr(n),\; n>1.
\end{matrix}
$$
The solutions are:
$$
\begin{matrix}
\mu=3-n, &\lambda=-1&\text{for  }\fns(n)\\
\mu=\frac{3}{2}, &\lambda=-\frac{1}{2}&\text{for  }\fr(1)\\
\mu=4-n, &\lambda=-1&\text{for  }\fr(n),\; n>1.
\end{matrix}
$$

\ssec{2.5.  The KdV hierarchies associated with the Schr\"odinger 
operators} Let $L_r$ be the Schr\"odinger operator of order $r$, see 
sec.  2.3.  Define the KdV-type equation as the following Lax $L$-$A$ 
pairs:
$$
D_\cT(L)=[L, A_k], \text{ where } A_k=(L^{k/r})_+ \text{ for } 
k\not\equiv r\pmod r  \eqno{(2.5.1)}
$$
and where 
$$
D_\cT=\left\{\begin{matrix}\frac{d}{dx}&\text{ if
$p(A_k)=\ev$}\\
\frac{\partial}{\partial \tau}+\tau\frac{\partial}{\partial x}&\text{ if
$p(A_k)=\bar{1}$}.\end{matrix}\right.
$$
Here the subscript $+$ singles out the differential part of the
pseudodifferential operator. For complex $k$ the differential part is not
well-defined and we shall proceed, mutatis mutandis, as 
Khesin--Malikov, namely, by setting
$$
D_\cT(L)=[L, A_k], \text{ where } A_k=(L^{1/r})^{k} \text{ for } 
k\in\Cee.  \eqno{(2.5.2)}
$$

\ssec{Pseudodiferential operators on the supercircle} Let $V$ be a
superspace. For $\theta=(\theta_1, \dots ,
\theta_n)$ set
$$
\begin{matrix}
V[x, \theta]=V\otimes \Kee [x, \theta];\; V[x^{-1}, x, \theta]=V\otimes
\Kee [x^{-1}, x, \theta];\\
V[[x^{-1}, \theta]]=V\otimes \Kee [[x^{-1},
\theta]];\\
V((x, \theta))=V\otimes \Kee [[x^{-1}]][x, \theta].\end{matrix}
$$
We call $V((x, \theta))x^\lambda$ the space of {\it pseudodifferential
symbols}. Usually, $V$ is a Lie (super)algebra. Such symbols correspond to
pseudodifferential operators (pdo) of the form
$$
\sum\limits_{i=-\infty}^n\sum\limits_{k_{0}+\dots+k_{n}=i}
a_i(\partial_x)^{k_{0}}\theta_1^{k_{1}}\dots \theta_n^{k_{n}},
$$
Here $k_i=0$ or 1 for $i>0$ and $a_i(x, \theta)\in V$. This is clear.

For any $P=\sum\limits_{i\leq m} P_ix ^i\theta_0^k\theta^j\in V((x, 
\theta))$ we call $P_{+}=\sum\limits_{i, j, k\geq 0} P_ix 
^i\theta_0^k\theta^j$ the {\it differential part} of $P$ and 
$P_{-}=\sum\limits_{i, k< 0} P_ix ^i\theta_0^k\theta^j$ the {\it 
integral} part of $P$.

The space $\Psi DO$ of pdos is, clearly, the left module over the 
algebra $\cF$ of functions.  Define the left $\Psi DO$-action on $\cF$ 
from the Leibniz formula thus making $\Psi DO$ into a superalgebra.

Define the involution in the superalgebra $\Psi DO$ setting
$$
(a(t, \theta)D^i\tilde D^j)^*=(-1)^{jip(\tilde D)p(D)}\tilde D^jD^ia^*(x,
\theta).
$$

The following fact is somewhat unexpected.  If $D$ is an odd 
differential operator, then $D^2$ is well-defined as $\frac{1}{2}[D, 
D]$.  Hence, we can consider the set $V((x, \theta))x^\lambda$ for an 
{\it odd} $x$!  Therefore, there are two types of pdos: {\it contact} 
ones, when $D^2\neq 0$ for odd $D$'s and general ones, when all odd 
$D$'s are nilpotent.

\begin{rem*}{Conjecture} There exists a residue for all distinguished
dimensions, i.e. for the contact type pdos in dimensions $1|n$ for $n\leq 4$
and for the general pdos in dimensions $1|n$ for $n\leq 2$.
\end{rem*}

So far, however, the residue was defined only for contact type pdos, of
$\fk^L$ type, and only for $n=1$ at that. 

Extend [MR] and define the {\it residue} of $P=\sum\limits_{i\leq m} 
P_ix ^i\theta_0^k\theta^j\in V((x, \theta_0, \theta))$ for $n=1$.  We 
can do it thanks to the following exceptional property of $\fk^L(1|1)$ 
and $\fk^M(1|1)$.  Indeed, over $\fk^L(1|1)$, the volume form \lq\lq 
$dt\pder{\theta}$" is, more or less, $d\theta$: consider the quotient 
$\Omega^1/\cF\alpha$, where $\alpha$ is the contact form preserved by 
$\fk^L(1|1)$; similarly, over $\fk^M(1|1)$, the transformation rules 
of $dt\pder{\theta}$" and $\tilde\alpha$, where $\tilde\alpha$ is the 
contact form preserved by $\fk^M(1|1)$, are identical.  
Therefore, define the residue by the formula
$$
\Res~P= \text{ coefficient of }\frac{\theta}{x}\text{
in the expansion of } P_{-1}.
$$

\begin{rem*}{Remark} Manin and Radul [MR] considered the
Kadomtsev--Petviashvili hierarchy associated with $\fns$, i.e., for
$D=K_\theta$. The formula for the residue allows one to directly generalize
their result and construct a simialr hierarchy associated with $\fr$, i.e.,
for $D=\tilde K_\theta$.
\end{rem*}

This new phenomenon --- an invertible odd symbol --- doubles the old 
picture: let $\theta_0$ be the symbol of $D$, and let $x$ be the symbol 
of the differential operator $D^2$.  We see that the case of the odd 
$D$ reduces to either $V((x, \theta_0, \theta))x^\lambda$ or $V((x, 
\theta_0^{-1}, \theta))x^\lambda$.  This is in accordance with the 
fact that every irreducible finite dimensional representation of 
$\fosp(1|2)$ is glued of two representations of $\fsl(2)$.

\end{document}